\documentclass{appolb}

\usepackage{epsfig}
\def\lsim{\mathrel{\raise.3ex\hbox{$<$\kern-.75em\lower1ex\hbox{$\sim$}}}}
\def\gsim{\mathrel{\raise.3ex\hbox{$>$\kern-.75em\lower1ex\hbox{$\sim$}}}}
\def\ga{\mathrel{\raise.3ex\hbox{$>$\kern-.75em\lower1ex\hbox{$\sim$}}}}
\def\la{\mathrel{\raise.3ex\hbox{$<$\kern-.75em\lower1ex\hbox{$\sim$}}}}
\begin{document}
\pagestyle{plain}
\title{$H^\pm \to W^\pm Z$, $H^\pm \to W^\pm \gamma$ in the MSSM 
\thanks{Presented at XXVII PHYSICS IN COLLISION - Annecy, France, 26 - 29 June 2007.}}
%==================
\author{A. Arhrib$^{1,2}$~, 
R. Benbrik$^{3,4}$~
and M. Chabab$^{4}$~
\address{
$^{1}$ Facult\'e des Sciences et Techniques B.P 416 Tanger, Morocco.\\
$^{2}$ National Central University, Physics Department, 
Chung-Li, Taiwan, R.O.C\\
$^{3}$ ChungYuan Christian University, Physics Department, 
Chung-Li,Taiwan,R.O.C\\
$^{4}$ LPHEA, Facult\'e des Sciences-Semlalia,
             B.P. 2390 Marrakesh, Morocco.}}
\maketitle
\vspace{-1cm}
\begin{abstract}
We study the complete one loop contribution to
$H^\pm\to W^\pm Z$ and $H^\pm \to W^\pm \gamma$ in the Minimal 
Supersymmetric Standard Model (MSSM). 
\end{abstract}
\PACS{11.15.Ex,11.30.Pb,14.70.-e}
At the LHC, the detection of light charged Higgs boson with
$m_{H^\pm}\la m_t$ is straightforward from top production 
followed by  the decay $t\to bH^+$.
Such light charged Higgs ($m_{H^\pm}\la m_t$) can be detected also 
for any $\tan\beta$ in
the $\tau\nu$ decay  which is indeed the dominant decay mode.
However, for heavy charged Higgs masses $m_{H^\pm}\ga m_t$
which decay predominantly to $t\bar{b}$, 
the search is rather difficult due to large irreducible and reducible
backgrounds associated with $H^+\to t\bar{b}$ decay.
However, it has been demonstrated  that the 
 $H^+\to t\bar{b}$ signature can lead to a visible signal at LHC
provided that the charged Higgs mass below 600 GeV and 
$\tan\beta$ is either below $\la 1.5$ or above $\ga 40$.
An other alternative discovery channel for heavy charged Higgs is  
 $H^\pm \to W^\pm h^0$, followed by the dominant decay of $h^0$ to $b\bar{b}$. 
This channel could lead to charged Higgs discovery
only for low $\tan\beta$  where the branching ratio of $H^\pm \to
W^\pm h^0$ is sizeable.

In MSSM, at tree level, the couplings $H^\pm \to W^\pm \gamma$, 
 $H^\pm \to W^\pm Z $ are absent. Therefore, 
decays modes like $H^\pm \to W^\pm \gamma $, 
$H^\pm \to W^\pm Z $ are mediated at one loop level and then are
expected to be loop suppressed. Moreover, those channels have
a very clear signature and might emerge easily at future colliders.
For instance, if $H^\pm \to W^\pm Z $ is enhanced enough, this decay 
may lead to three leptons final state if both W and Z decay leptonically
and that would be the corresponding golden mode for charged Higgs boson.

We have evaluated the one-loop induced process
$H^\pm \to  W^\pm V$ in the 'tHooft-Feynman gauge using
dimensional regularization. We also use the on-shell 
renormalization scheme for Higgs sector. (see \cite{abc} for more details).
Form factors of $H^\pm \to W^\pm \gamma$ are constrained by 
electromagnetic gauge invariance while those of $H^\pm \to W^\pm Z$ are not.
In this respect, $H^\pm \to W^\pm Z$ will be more enhanced 
by SUSY contributions than $H^\pm \to W^\pm \gamma$.\\
%
%======================================================
%%%%%%%%%%%%%%%%%%%%%%%%%%%%%%%%
\begin{figure}[h]
\begin{center}
\epsfig{file=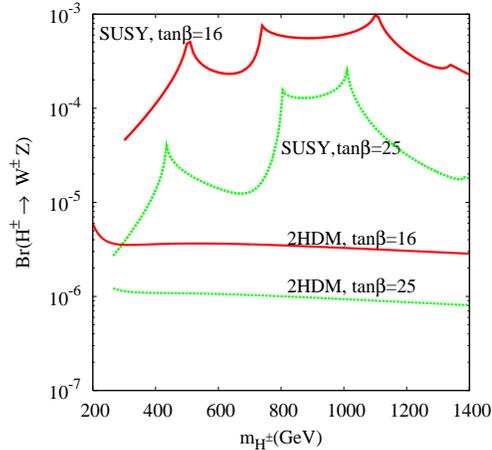,width=8.7cm,angle=0}
\caption{
{\small Br($H^{\pm}\to W^{\pm}Z$) as a function of 
$m_{H^{\pm}}$  in  the MSSM and 2HDM for $M_{SUSY},M_2,\mu=500,175,
-1400$ GeV and $A_{t,b,\tau}= -\mu$  for various values of $\tan\beta$}}
\end{center}
\end{figure}
Our main result is shown in Fig.1 where we have 
illustrated Br($H^{\pm}\to W^{\pm}Z$) as a function 
of charged Higgs mass. 
Numerically, the Br($H^\pm \to W^\pm \gamma$) 
never exceed $10^{-5}$ and will not be show here 
(see \cite{abc} for more details).
In Fig.1, we have shown both the
pure 2HDM and the full MSSM  contribution. It is clear that the 
2HDM contribution is rather small. Once we include the SUSY particles, 
we can see that the Branching fraction get enhanced and can reach $10^{-3}$. 
The source of this enhancement is mainly due 
to the presence of scalar fermion contribution in the loop which are
amplified by threshold effects from the opening of the decay 
$H^\pm \to \widetilde{t}_i\widetilde{b}_j^*$. It turns out that the 
contribution of charginos neutralinos loops does not enhance the 
Branching fraction significantly as compared to scalar fermions loops.
The plot also show that, the branching fraction is more important for 
intermediate $\tan\beta=16$ and is slightly reduced for larger $\tan\beta=25$.
\\
To conclude, those Branching ratios of the order $10^{-3}$ might provide 
an opportunity to search for a charged Higgs boson at the LHC 
through $H^\pm \to W^{\pm} Z$. The smallness of those branching ratios
may require high luminosity option as is already planned with SuperLHC.

\end{document}